# A DFT+U study of the segregation of Pt to the $CeO_{2-x}$ $\sum 3[1\bar{1}0]/(111)$ grain boundary

Zhixue Tian

*Department of Physics, Hebei Normal University, Shijiazhuang, Hebei 050024, China*

**Abstract**

Grain boundaries (GBs) can be used as traps for solute atoms and defects, and the interaction between segregants and GBs is crucial for understanding the properties of nanocrystalline materials. In this study, we have systematically investigated the tendency of Pt to segregate as well as the interaction between Pt and oxygen vacancies at the $\sum 3$ (111) GB of ceria ($CeO_2$). The Pt atom has a stronger tendency to segregate to the $\sum 3$ (111) GB than to the (111) and (110) free-standing surfaces, but the tendency is weaker than to the (100) free-standing surface. Mechanic contributions (lattice distortion) play a dominant role in the strong tendency of Pt to segregate. At the Pt-segregated-GB (Pt@GB), oxygen vacancies prefer to form spontaneously near Pt in the GB region. However, at the pristine GB (no Pt and no vacancies), oxygen vacancies form only under O-poor conditions. Thus, Pt segregation to the GB promotes the formation of oxygen vacancies, and their strong interactions enhance the interfacial cohesion. We propose that GBs fabricated close to the surfaces of nanocrystalline ceria can trap Pt from the bulk or other types of surface, resulting in the suppression of the accumulation of Pt on the surface under redox reactions, especially under O-poor conditions.

# 1. INTRODUCTION

Grain boundaries (GBs) are important planar defects, ubiquitous in all polycrystalline materials, and are interfaces formed naturally by numerous single-crystalline grains which are bonded to one another.[1] Due to symmetry breaking and the presence of excess volume at the boundary, GBs can be used as traps for some solute atoms and point defects inside the bulk. The segregation of solute atoms at GBs and the interaction between the intercrystalline defects and atomic defects impact the local structure and chemistry, and thus govern the properties of many materials.[2,3,4,5] Additionally, in nanocrystalline ceramic materials, GBs or interfaces and the defects associated with GBs are also major factors contributing to the size dependency of chemical [6,7,8,9], electrical[10], ionic[11], and magnetic[12,13] properties. Therefore, understanding the character of GBs and the interactions of defects in GBs is crucial for designing and optimizing nanostructured materials.

Ceria ($CeO_2$) has attracted considerable scientific interest because of its applications in many areas, including chemical catalysis, solid oxide fuel cells, and oxygen storage devices. As a support of the three way catalyst (TWC) which converts internal combustion exhaust emissions into harmless gases, $CeO_2$ can stabilize the precious metal (Pt) on its surfaces, avoiding the growth and aggregation of Pt particles, and thus maintaining and improving the activity of the TWC. By lowering the surface energy of Pt nanoparticles and trapping Pt in an atomically dispersed state, $CeO_2$ can effectively inhibit the sintering of Pt particles.[14] Many experimental and theoretical investigations have reported the segregation, bonding, and migration behaviors of the doped Pt ions and Pt clusters on the $CeO_2$ surface with and without oxygen vacancies.[15,16,17,18,19,20,21,22,23,24] Indeed, $CeO_2$ is widely considered to be not merely a support for the precious metals but also an energetic participant in the catalytic process.[21,25,26] This is because $CeO_2$ is able to contain numerous oxygen vacancies and Ce ions, leading to excellent oxygen mobility and oxygen storage capacity of the materials.[27,28] As a result, the catalytic activity tends to occur in limited regions where the precious metal, $CeO_2$, and the reactants are all present simultaneously.

Therefore, in order to improve the catalytic activity, one approach is to increase the number of such catalytic centers in which Pt particles are dispersed on the $CeO_2$ surface in very small clusters. This means that Pt particles must be anchored to the support, in such a manner as to inhibit diffusion and aggregation on the surface. At present, a new generation of precious metal ionic catalysts[20,21] is being explored to promote the activity of $CeO_2$ catalysts by greatly reducing the size of the precious particles. In such materials, Pt is ionically dispersed into the $CeO_2$ host matrix, and strongly anchored on the surface via segregation in an oxidative treatment. These new catalysts exhibit more activity than previous ones in which precious metals were directly dispersed on oxide supports and were thus prone to coalescing[20,21]. However a solid solution of Pt obtained by replacing Ce atoms in the $CeO_2$ bulk with Pt atoms has been found to be unsuitable, being dominated by the smaller ionic radii of $Pt^{2+}$ or $Pt^{4+}$ oxidation states.[14] In $CeO_2$, cerium atoms are surrounded by an eight-fold coordination of the O atoms, Pt cations, however, prefer six-fold coordination with O atoms, thus leading to the segregation of Pt cations to the $CeO_2$ surfaces.[14]

In addition to the surfaces, $CeO_2$ crystals also contain GBs which are typical planar defects as well, which can also trap solid solutes and vacancies. The typical $CeO_2$ model GBs, such as $\sum 3 [1\bar{1}0]/(111)$, $\sum 5 [001]/(210)$, $\sum 9 [110]/(211)$, $\sum 11 [110]/(332)$, and $\sum 13 [001]/(510)$ GBs, have been made experimentally using a bicrystal technique.[29] The local GB structures have been identified at the atomic scale using scanning transmission electron microscopy (STEM) combined with calculations based on density functional theory (DFT). Oxygen vacancies, which are common point defects, often result in a variety of additional functionalities, and have thus been widely studied not only in the bulk case but also for GBs.

On the basis of previous studies, it is known that the formation of oxygen vacancies depends strongly on the atomic structure, largely because of the distortions of the local structure of the GB.[30] According to both STEM and EELS (electron energy-loss spectroscopy) studies, oxygen vacancies appear spontaneously in the $CeO_2$ $\sum 5$ [001]/(210) GB, and stabilize the GB structure. On the other hand, the $CeO_2$ $\sum 3[1\bar{1}0]/(111)$ GB, whose structural distortion is about 8.5 nm$^{-2}$, much smaller than that (17.8 nm$^{-2}$) of $\sum 5[001]/(210)$ GB [29], fails to promote the formation of oxygen vacancies, and maintains the oxygen stoichiometry.[29,31] However, oxygen vacancies tend to segregate to the $CeO_2$ $\sum 3[1\bar{1}0]/(111)$ GB from the bulk[32], especially in the case of Mn- and Gd-doped GBs[33]. The segregation of oxygen vacancies near GBs affects the oxygen storage capability of nanostructured $CeO_2$ and provides positions for high concentration of oxygen vacancies crucial for chemical catalysts in the automotive industry. Thus, the complex composed by metal-solute and oxygen vacancies can influence the properties of $CeO_2$ GBs, and deserves further investigation.

Though Pt is the most common metal-solute used as a catalyst in nanocrystalline $CeO_2$, its segregation and interactions with oxygen vacancies at the GB are as yet unclear. Since the behavior of Pt atoms could be one of the key points in understanding the dispersion of Pt on the $CeO_2$ surface, we have investigated Pt segregation and consequent effects on the formation of oxygen vacancies at the $CeO_2$ $\sum 3[1\bar{1}0]/(111)$ GB. Compared with the $\sum 5$ [001]/(210) GB, the $\sum 3$ $[1\bar{1}0]/(111)$ GB has less structural distortion and a lower density of three-fold groupings of O atoms at the GB plane, and thus its interfacial formation energy is lower than that of the $\sum 5$ [001]/(210) GB. For the same reason as mentioned in Ref. (32), we have systematically calculated from first-principles, the Pt segregation properties at the $CeO_2$ $\sum 3[1\bar{1}0]/(111)$ GB with and without an oxygen vacancy. For comparison, we have also investigated the states of Pt in the bulk and on three kinds of surfaces ((111), (110), and (100)) of $CeO_2$. As expected, Pt has a strong tendency to accumulate at the $\sum 3[1\bar{1}0]/(111)$ GB, even stronger than that at the (111) and (110) surfaces, but weaker than at the (100) surface. Moreover, oxygen vacancies prefer to form in the

Pt-segregated-GB (Pt@GB). Thus, Pt segregation can facilitate the spontaneous formation of oxygen vacancies, indicating the strong attraction between them.

The remainder of this Article is organized as follows: In section II, we show the technical details of our computational method. In sections III and IV, we present and discuss the properties of Pt at the $CeO_2$ $\sum 3[1\bar{1}0]/(111)$ GB, as well as the formation of oxygen vacancies with and without the segregated Pt. Finally, we give a brief summary of our results in section V.

## 2. METHODOLOGY

Total energy calculations were performed using the Vienna ab initio simulation package (VASP)[34,35] based on density functional theory (DFT). The projector augmented wave (PAW) method[36] was used to describe the electron-ion interaction, and the calculation of the exchange correlation between electrons used the generalized gradient approximation (GGA) in the Perdew-Burke-Ernzerhof (PBE) form[37] coupled with the Hubbard on-site Columbic correction (GGA+U)[38,39]. The value $U_{eff}(Ce) = 6$ eV was set to capture the localized state of the electrons into 4$f$ orbitals. To evaluate the effect induced by the value of $U_{eff}(Pt)$ on the properties of Pt at the $\sum 3[1\bar{1}0]/(111)$ GB, we calculated the segregation energies of Pt at the GBs (with and without an oxygen vacancy) and at the surfaces for $U_{eff}(Pt) = 0$ and 7.5 eV[40], with $U_{eff}(Ce) = 6$ eV fixed for Ce 4$f$. The Ce 5$s$, 5$p$, 5$d$, 4$f$ and 6$s$ electrons, the O 2$s$ and 2$p$, and the Pt 5$p$, 5$d$, and 6$s$ electrons were treated as valence electrons. Spin-polarization was employed in all calculations. The cutoff energy for the plane wave basis set was 400 eV. We used a (2×4×1) $k$-mesh[41] within the Monkhorst-Pack scheme for all GB systems. All the atomic structures were fully relaxed until the residual force on each atom was less than 0.01 eV/Å. The calculated lattice constant and band gap for the bulk ceria were about 5.493 Å and 2.40 eV, respectively. Our results are consistent with a previous DFT+U calculation by Shi $et$ $al$[42], but deviate slightly from the experimental values of 5.41 Å and 3.0 eV for lattice constant and band gap.[43,44,45].

**Figure 1. A side view along the $[1\bar{1}0]$ axis of a tilt $CeO_2$ ∑3$[1\bar{1}0]$/(111) GB. The violet and red balls represent the Ce and O atoms, respectively. The region between the two dashed lines indicates the ∑3 GB of $CeO_2$. GB1 and GB2 represent the Ce sites for Pt segregation. GB1 is at the GB region, while GB2 is located at the atomic layer between the bulk and GB.**

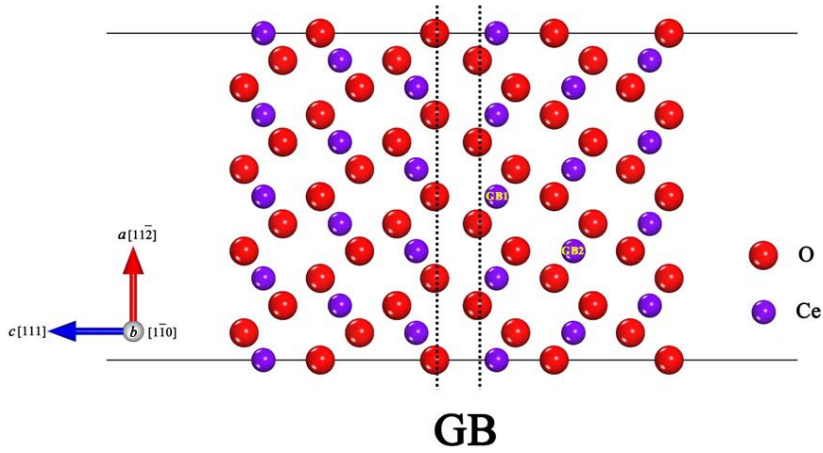

A tilt ∑3$[1\bar{1}0]$/(111) GB composed of two identical grains (each one containing three stoichiometric $CeO_2$ (111) trilayers, i.e., nine atomic layers) is shown Figure 1. We used slabs terminating with a vacuum region of about 18 Å to simulate GBs. We modeled the GB configuration by mirroring and shifting the (111) plane of each grain as in Ref. (31) and (32). To ensure a sufficient size in the *xy*-dimension, we calculated the Pt-Pt interactions in a (2×2×2) $CeO_2$ bulk. When the distance of the Pt-Pt pair in the $[11\bar{2}]$ direction was about 6.728 Å ($\sqrt{6}a/2$, where *a* is the lattice constant for bulk $CeO_2$), the two Pt atoms attract each other strongly. However, when the two Pt atoms were separated in $[1\bar{1}0]$ direction by 7.768 Å ($\sqrt{2}a$), there was almost no interaction between them. Therefore, we constructed a $(\sqrt{6}\times\sqrt{2})a$ unit cell in the *xy*-2-dimensional plane to avoid lateral interactions between two periodic Pt atoms.

This unit cell is large enough to study the properties of Pt at the ∑3 (111) GB of $CeO_2$ with and without an oxygen vacancy. A pristine ∑3 (111) GB of $CeO_2$ as used in our calculations thus contained 144 atoms.

In the geometry optimizations, we fixed three terminal atomic-layers (namely, one stoichiometric $CeO_2$ (111) trilayer) in each grain at the bulk position. Each Ce atom in the GB region had seven nearest neighbor O atoms all with the same Ce-O bond length of about 2.377 Å. The (111) surface was modeled using nine atomic layers. During the energy minimization, we fixed three terminal layers. We also calculated the solution energies of Pt on the $CeO_2$ (110) and (100) surfaces to compare the segregation tendency with that at the GB. Thus, we used the same models for $CeO_2$ (110) and (100) surfaces following the description in Ref. (46).

The solution energy of a substitutional Pt in the host $CeO_2$ systems (bulk, GB, and surfaces) with and without an oxygen vacancy is defined generally as:

$$E_{Pt}^{sol} = E_{tot}(Pt@host) - E_{tot}(host) - \mu_{Pt} + \mu_{Ce}, \quad (1)$$

where $E_{tot}(Pt@host)$ and $E_{tot}(host)$ are the total energies of the systems with and without a Pt atom, and $\mu_{Pt}$ and $\mu_{Ce}$ denote the chemical potential for Pt and Ce, respectively. The segregation energy of Pt at the GB or on surfaces can then be written as:

$$E_{seg}^{Pt} = E_{sol}(GB/surf.) - E_{sol}(bulk) \quad (2)$$

where $E_{sol}(GB/surf.)$ is the solution energy of Pt at/near the GB or on the (111) surface. $E_{sol}(bulk)$ is the solution energy of Pt when it substitutes for a Ce atom in a (2×2×2) supercell of $CeO_2$.

The formation energy of a charged oxygen vacancy at the GB was defined as:

$$\Delta E_{V_O}^{f} = E_{tot}(GB\text{-}V_O, q) - E_{tot}(GB, 0) - n_{V_O}\mu_O + q(\mu_e + \varepsilon_v) \quad (3)$$

in which $E_{tot}(\text{GB-V}_O, q)$ and $E_{tot}(\text{GB}, 0)$ are the total energies of the ∑3 (111) GBs (pristine or Pt-segregated GB (Pt@GB)) with and without oxygen vacancies. $n_{V_O}$ is the number of oxygen vacancies. In the present work, only one vacancy was considered at the GB, and, thus $n_{V_O} = 1$. $\mu_O$ is the chemical potential of the removed O atom, and is equal to the total energy per atom in its elemental crystal. The range of the oxygen chemical potential is from -9.14 eV (O-poor) to -4.93 eV (O-rich), which was determined using two thermodynamic limits: (1) the limit against decomposition of CeO$_2$ into an *fcc* crystal of Ce and O$_2$ gas, and (2) the stoichiometric limit of CeO$_2$. Finally, $q$ is the charge state of the oxygen vacancy, and $\mu_e$ is the electron chemical potential with respect to the energy level ($\varepsilon_v$) of the host valence band maximum (VBM). Therefore, $\mu_e$ can vary from zero to the value of the host band-gap.

The charge density difference induced by the segregated Pt at the GB with and without an oxygen vacancy is calculated using:
$$\Delta\rho = \rho(\text{Pt@GB-}n\text{V}_O) - \rho(\text{V}_{Ce}\text{@GB-}n\text{V}_O) - \rho(\text{Pt})$$
where the first item represents the charge density of the Pt@GB with ($n$ = 1) and without ($n$ = 0) an oxygen vacancy; the middle item gives the charge density of the GB-$n$V$_O$ with a vacancy of Ce; the last item is the charge density for a Pt monolayer, which is located at the same site as the Ce vacancy.

## 3. RESULTS AND DISCUSSIONS

### 1. Pt segregation on the ∑3[1$\bar{1}$0]/(111) GB.

With respect to Pt inside the bulk, we traced a series of solution energy changes for Pt: first for the GB, and then for the (111), (110), and (100) surfaces. To understand the effect of the Coulomb repulsion term, $U_{eff}$, on Pt segregation, the relative solution energies (segregation energies) of Pt to the ∑3[1$\bar{1}$0]/(111)GB and the surfaces for

$U_{eff}(Pt) = 0$ and 7.5 eV, with $U_{eff}(Ce) = 6$ eV, were calculated. The calculated results are presented in Fig. 2. It was found that Pt had a strong tendency to segregate to the GB and surfaces whatever the value of $U_{eff}$. The (100) surface was the most favorable site for Pt, followed by the $\Sigma 3[1\bar{1}0]/(111)$ GB. Pt had similar tendencies to segregate to the (111) and (110) surfaces. The Coulomb repulsion of the Pt 5d orbital raises the relative energies for Pt segregating to the GB, (111) and (110) surfaces by 0.20~0.40 eV, but reduces the energy for Pt segregating to the (100) surface by 0.4 eV. In the [111] direction, the $\Sigma 3[1\bar{1}0]/(111)$ GB was more favorable than the site on the (111) surface, with energy differences of about 0.57 and 0.61 eV for $U_{eff}(Pt) = 0$ and 7.5 eV, respectively. Based on our previous investigations and the references therein[47], this implies that segregated Pt can improve the GB cohesion. In order to verify the reliability of our result, we calculated the total energy difference between Pt at the GB site and on the top site of a (111) surface in an expanded supercell that consisted of a $(\sqrt{6} \times 2\sqrt{2})a$ unit cell in an *xy*-2-dimensional plane along with four stoichiometric $CeO_2$ (111) trilayers in each grain, for a total 384 atoms. The calculated total energy difference of about -0.33 eV was only a little greater than that calculated in the smaller systems mentioned above. Thus, our computational models are large enough to provide reliable results. In addition, the magnetic properties and structures of Pd-doped-GBs (with and without oxygen vacancies) and surfaces are almost the same, except for a few slight changes in the Pt-O bond length. Therefore, the segregation properties of Pt in GB can be reasonably described using the GGA+U method with $U_{eff}(Ce) = 6$ eV and $U_{eff}(Pt) = 0$ eV.

**Figure 2. The calculated relative solution energies of Pt as a function of its position. The solution energy of Pt inside the bulk is considered as the reference state, i.e. the zero point. The blue and red segments represent the calculated relative energies using $U_{eff}(Pt) = 0$ and 7.5 eV, respectively. The blue line traces**

**the segregation from the bulk to the (111) surface and the ∑3 (111) GB.**

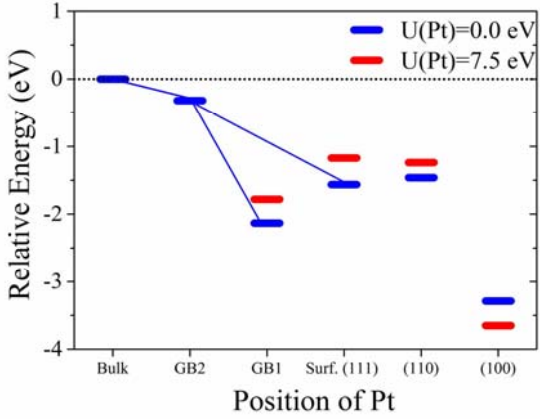

To identify explicitly the main contributions to Pt segregation, we separated the solution energy of Pt in each $CeO_2$ system into the mechanical and chemical contributions. We took the mechanical part to be the energy change due to the lattice relaxation. Therefore, the mechanical contribution is defined as:

$$\Delta E_{mec}^{Pt} = E_{tot}^{rel.}(Pt@host) - E_{tot}^{unrel.}(Pt@host) \tag{4}$$

where $E_{tot}^{rel.}(Pt@host)$ and $E_{tot}^{unrel.}(Pt@host)$ are the total energies of the Pt@host with and without relaxation. It should be pointed out that Pt is doped in a fully relaxed $CeO_2$ system, and thus the energy change associated with the lattice distortion derives from having the Pt atom instead of the Ce at a particular site. In addition, during the energy minimization of $Pt@CeO_2$ systems, every ionic step is fully self-consistent.

The other is the chemical contribution which is calculated as:

$$\Delta E_{chem}^{Pt} = E_{tot}^{unrel.}(Pt@host) - E_{tot}^{rel.}(host) + E_{bulk}(Ce) - E_{bulk}(Pt) \tag{5}$$

where $E_{tot}^{rel.}(host)$, $E_{bulk}(Ce)$, and $E_{bulk}(Pt)$ represent the total energy of the pristine relaxed $CeO_2$ system, and total energies of Ce and Pt atoms in their bulk. Figure 3 shows the solution energies and the contributions of Pt in the bulk, ∑3 (111) GB, and three surfaces ((111), (110), and (100)) of $CeO_2$. We found that the solution energies of Pt are largely due to the chemical part. It seems that as the number of nearest O atoms decreases, the chemical influence on the solution energy of Pt decreases.

However, due to the similar magnitude of the chemical contributions for all five systems, the relative solution energy with respect to Pt inside the bulk, i.e. the segregation tendency, relies mainly on the mechanical contribution. That is, the greater the lattice distortion induced by Pt replacing a Ce atom, the stronger the segregation tendency of Pt. On the (111) surface, the energy release associated with the surface relaxation induced by the Pt atom is about -5.036 eV, indicating the stronger segregation tendency than for the other four systems. The next one is Pt@∑3 (111) GB with the mechanical contribution being about -2.168 eV.

**Figure 3. The calculated solution energy along with the mechanical and chemical contributions for Pt in the bulk, at the ∑3 (111) GB and the surfaces. The mechanical contribution is defined as the total energy release during the lattice relaxation. The chemical contribution is associated with the energy change due to the replacement of Ce by Pt.**

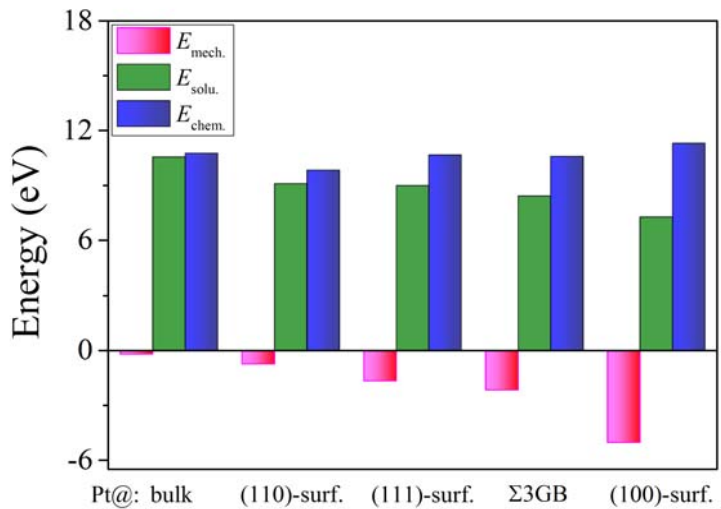

We next discuss the segregation properties of Pt in the [111] direction. In the $CeO_2$ bulk, the Pt substituent is surrounded by eight oxygen anions, all with the same Pt-O bond length of about 2.30 Å. Moreover, from the projected density of states (PDOS) of Pt, we can note a significant spin-up state appearing at the Fermi level, indicating a $Pt^{2+}$ oxidation state inside the bulk, which is qualitatively consistent with previous experimental and theoretical studies. When Pt replaces the Ce atom at the ∑3 (111) GB and the (111) surface, however, there are only six oxygen anions around it. In the case of Pt@GB, Pt retains its six-fold coordination, but the Pt-O bond lengths are not identical. Three Pt-O bonds with O atoms near the GB are dramatically decreased to 2.02 ± 0.01 Å, while the other three Pt-O bonds with the O atom near the bulk are 2.19 Å long. On the (111) surface, three Pt-O bonds with O exposed to the surface have an average length of about 2.20 Å, and the others are about 2.09 Å. The PDOSs of Pt at the GB and the surface look very similar near the Fermi level. Both of them have a band gap of about 1.50 eV, but no spin-polarized state. However, the energy level of the binding state for Pt at the GB is slightly lower than that for Pt on the (111) surface, leading to stronger Pt-O binding at the ∑3 (111) GB.

**Figure 4. Projected DOS (PDOS) plotted in (a)-(c) are for the Pt atom in the bulk, ∑3 (111) GB, and the (111) surface. The Fermi level is set to zero. The spin-density distribution induced by Pt in the bulk is presented in (d). The isosurface value is 0.005 e/a.u.$^3$. The octahedrons comprised of a Pt atom and six O atoms which are the nearest neighbors for the segregated-Pt at the GB and the (111) surface are shown in panels (e) and (f), respectively. The bond lengths for the Pt-O bonds are also labeled.**

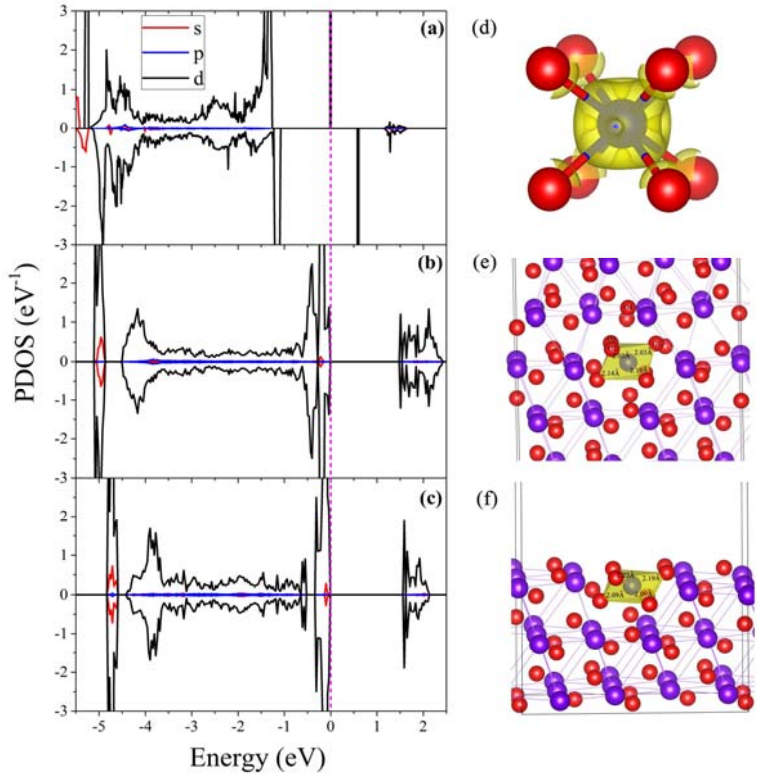

**2. The interaction of Pt and oxygen vacancies on the CeO$_2$ ∑3[1$\bar{1}$0]/(111) GB.**

It has been found that oxygen vacancies have a significant tendency to segregate onto the CeO$_2$ ∑3 (111) GB, as mentioned in Ref. (32). In the present study, we compared the stability of an oxygen vacancy in the bulk and at the ∑3 (111) GB of CeO$_2$, and found that the ∑3 (111) GB is much more favorable than the bulk for an oxygen vacancy with the energy difference being about 2.37 eV. For the ∑3 (111) GB with Pt also segregating, we examined the stability of four possible sites for the oxygen vacancy by removing an O atom near the center of the GB region (See Fig. 4e). As mentioned above, Pt is surrounded by six O atoms, including three O atoms at the GB. Although these three Pt-O$_{1/2/3}$ bonds have similar bond lengths and charge density distributions, the energies needed to break the bonds are rather different. That is, the formation energies of these oxygen vacancies are different. Besides the nearest neighbors, we also examined the stability of the vacancy at the second nearest neighbor site of Pt. Using Eq. (3), we calculated the formation energy of an oxygen vacancy with and without a segregated Pt under the O-rich ($\mu_O$ = -4.93 eV) and O-poor ($\mu_O$ = -9.14 eV) conditions, respectively.

**Figure 5.** Formation energies of oxygen vacancies at different GB sites with and without a segregated-Pt as a function of the Fermi level under the O-rich and O-poor conditions. The charge states are shown for each line segment. The symbols indicate the inflection points for the charge state changes of oxygen vacancies.

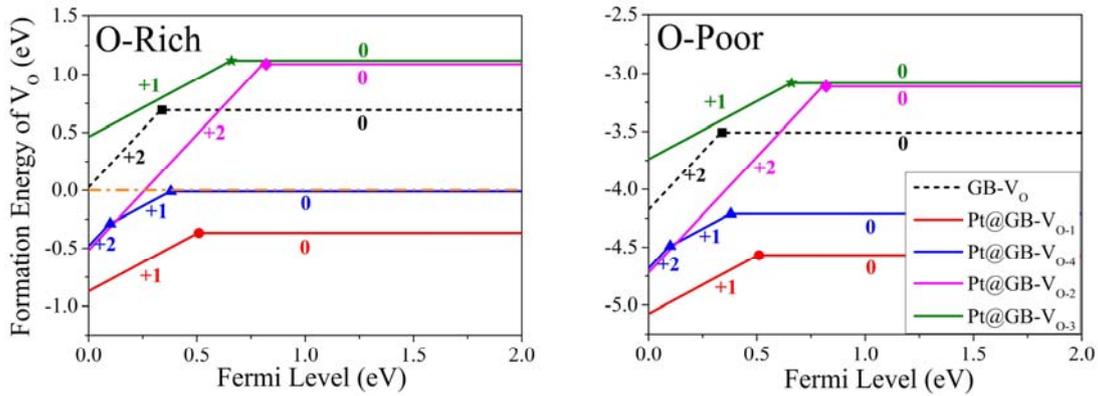

**Figure 6.** The calculated formation energies of an oxygen vacancy at the $V_{O-1}$ site with (b) and without (a) the segregation of Pt as functions of the chemical potential of O and an electron.

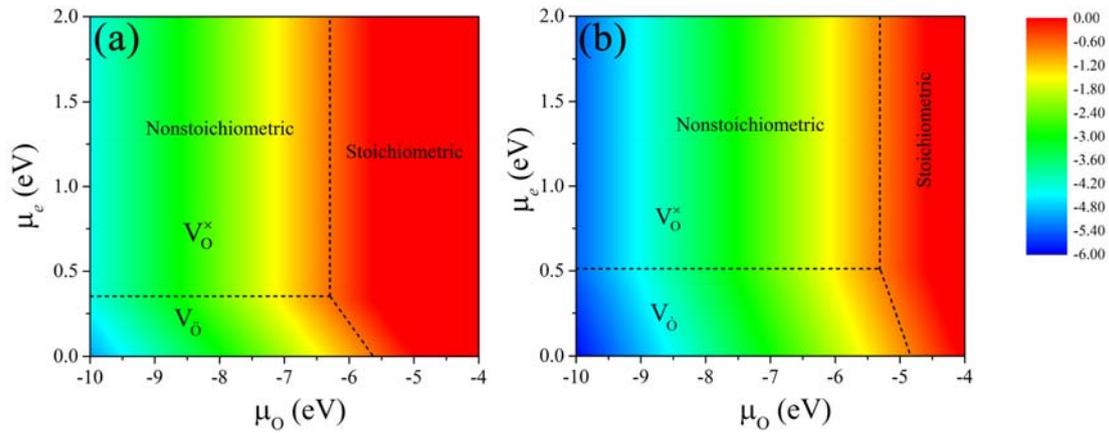

Figure 5 shows the calculated formation energy of an oxygen vacancy at the pristine GB and at different sites in this GB, Pt@GB, as a function of the Fermi level, under both O-poor and O-rich conditions. Under the O-poor conditions, all $V_O$s have negative formation energies over a wide range for the Fermi level, indicating that they can form spontaneously during crystal growing or later, especially in the Pt@GB system. Thus, the concentration of oxygen vacancies at the GB is expected to be very high under O-poor conditions. On the contrary, due to the greatly increased formation energies, no vacancies except for $V_{O-1}$ and $V_{O-4}$ form naturally in the Pt@GB under the O-rich condition, indicating a low concentration of oxygen vacancies. Given the difference between the GB with and without a Pt atom, the Pt@GB-$V_{O-1}$ (or Pt@GB-$V_{O-4}$) has a lower formation energy than the GB-$V_O$ over the whole range of Fermi levels, and they can even form spontaneously in the Pt@GB under the O-rich condition. Therefore, oxygen vacancies prefer to form at the Pt@GB rather than at a site without a Pt atom, indicating that the segregated-Pt and $V_{O-1}$ (or $V_{O-4}$) can attract each other strongly in the GB region. Additionally, Pt@GB-$V_{O-1}$ is energetically favored by about 0.5 eV over Pt@GB-$V_{O-4}$. The phase diagram as a function of the chemical potential of O and the electron for the formation energy of $V_{O-1}$ at the GB with and without a Pt atom are shown in Fig. 6.

The possible charge states and the relevant charge transition levels of the vacancies, which are very important for determining the defect states that actually occur, can also be seen from Fig. 5. The inflection points which are highlighted by colored symbols represent the transition points where the stable charge state of the $V_O$ changes. At the pristine GB, the formation energy of an oxygen vacancy shows a change of slope from $q = 2+$ to $q = 0$ (2+/0 transition) with the transition occurring about 0.31 eV above the valence band minimum (VBM). The same result can also be found in Fig.6a. Under O-poor conditions, vacancies with charges of 2+ ($V_O^{\bullet\bullet}$) and 0 ($V_O^x$) form naturally at their respective energy levels, because of their negative formation energies. $V_O^{\bullet\bullet}$ is dominant at a low Fermi level (p-type doping) and is located in the O-poor region, whereas $V_O^x$ is likely to form at a high Fermi level (n-type doping) and in an O-poor region. We did not observe an oxygen vacancy with a charge of 1+ ($V_O^{\bullet}$) which indicates that in our calculation, the $V_O^{\bullet}$ vacancy is unstable for any Fermi level in the GB without a Pt atom. There is thus no region in Fig. 6a identified with $V_O^{\bullet}$. This is a little bit different from the result in Ref. (32), in which $V_O^{\bullet}$ is stable, but in only a very limited region. The reason for the absence of $V_O^{\bullet}$ can be attributed to the negative-U[48] behavior which results in obvious differences in the lattice optimization between different charge states of oxygen vacancy. In the case of Pt@GB, the formation energies for $V_{O-1}$ and $V_{O-3}$ show one transition from $q$ = 1+ to $q$ = 0 with the transition levels at about 0.31 and 0.66 eV above the VBM. However, the formation energy of $V_{O-3}$ is greater than that of $V_O$ in the pure GB, especially for its neutral charge state, and thus it fails to form in the Pt@GB. From Fig.5, it may be seen that the $V_{O-1}$ vacancy forms naturally near the segregated-Pt GB, and is largely unaffected by external conditions. Note that the formation energy is dramatically lower, by as much as 1.0 eV, compared to that in the pristine GB. From Fig. 6b, the $V_O^{\bullet}$ vacancy is seen to occur at lower Fermi levels, and $V_O^{\times}$ prefers the higher Fermi levels. The transition level from $q$ = 1+ to $q$ = 0 is about 0.51 eV above the VBM. The next stable vacancy is $V_{O-4}$, which is located in the other grain facing the Pt atom. Its formation energy has two changes of slope, one is from $q$ = 2+ to $q$ = 1+ with the transition level at 0.1 eV, while the other is from $q$ = 1+ to $q$ = 0 with the transition level being 0.38 eV. At the Fermi level, the $V_{O-2}$ vacancy with a 2+ charge state has a lower formation energy just above the VBM ($0 < \varepsilon_F < 0.1 \text{eV}$), which is very similar to the formation energy of the $V_{O-4}$ vacancy in the 2+ state. As the Fermi level increases, its formation energy gradually increases even more than that of the $V_O$ in the pristine GB. For the neutral state appearing at a Fermi level of 0.82 eV, the formation energy of the $V_{O-2}$ vacancy is as large as that of the $V_{O-3}$ vacancy. Taken as a whole, one can see that the formation energies and transition energies of oxygen vacancies on the four nonequivalent O sites in the Pt@GB are very different due to their distinct environments surrounding the Pt atom. Pt@GB-$V_{O-1}$ is the most favorable configuration, followed by $V_{O-4}$ and $V_{O-2}^{\bullet\bullet}$ in a very low and limited energy region.

To find out if the Coulomb repulsion term $U_{\text{eff}}(\text{Pt})$ has any effect on the Pt-$V_O$

interaction, we also calculated the total energy of the Pt@GB-$V_{O-1}$ system with $U_{eff}(Pt) = 7.5$ eV and $U_{eff}(Ce) = 6$ eV. The difference in the formation energy of the oxygen vacancy is only 0.15 eV between the calculations for $U_{eff}(Pt) = 0$ and 7.5 eV. In addition, the magnetic properties and structures of Pt@GB-$V_{O-1}$ are almost the same with $U_{eff}(Pt) = 0$ as with 7.5 eV, except for slightly reduced Pt-O bond lengths. Therefore, the strong attraction between Pt and $V_O$ is unaffected by the value of $U_{eff}(Pt)$. We used $U_{eff}(Pt) = 0$ eV and $U_{eff}(Ce) = 6$ eV in the following calculations and discussions of the electronic properties

We propose that the stability of the vacancy-containing Pt@GB can be related to the coordination environment of the Pt atom. Similar to the case of a Pt atom in the Pt@GB without a oxygen vacancy, the Pt atom is surrounded by six O atoms as its nearest neighbors when a vacancy forms at the $V_{O-1}$ or the $V_{O-4}$ site (see Fig. 7b and 7d). In the case of the Pt@GB-$V_{O-4}$ system, the Pt-centered-octahedron looks similar to that in the original Pt@GB, except for the changes in the bond lengths and bond angles. The lengths of the Pt-O bonds which are in the GB region decrease by 1.5%, whereas the other three bonds away from the GB region are increased by 10.4%. Therefore, the interactions between Pt and the three O atoms farther away from the GB are weakened. When the $V_{O-1}$ vacancy appears adjacent to Pt, although Pt loses one nearest O atom, the Pt-centered-octahedron is still formed across the GB region, and the two grains are tightly connected by the octahedron. Thus, the Pt-$V_{O-1}$ interaction strengthens the interfacial bonding of the GB significantly. In contrast, Pt has four nearest O atoms in the cases of the $V_{O-2}$ and $V_{O-3}$ vacancies. However, due to the large difference in the lattice relaxation, the coordination number of Pt returns to six in the case of the 2+ charged state for $V_{O-2}$, which can explain why the $V_{O-2}^{\bullet\bullet}$ vacancy is stable only at low Fermi levels. The same conclusion is obtained in the analysis of the charge difference induced by Pt in the vacancy-containing GB systems as well. The results are presented in Fig. 7.

To facilitate comparison, the charge difference of Pt in the $CeO_2$ $\sum 3$ GB without vacancies is also plotted. On the basis of Bader-charge analysis[49], the Pt atom in Pt@GB loses a total of about 1.5 $e$, and transfers these electrons to its nearest neighbor O atoms. The charge accumulations on six Pt-O bonds are clear in Fig. 7. The transferred charge comes from the $t_{2g}$ orbitals of Pt. For the vacancy formation at the $V_{O-1}$ site in the Pt@GB, Pt transfers the same number of electrons from $e_g$ orbitals to its new surrounding O atoms. Thus we find a different octahedron, which is perpendicular to the plane of the GB. The accumulated charge across the GB enhances the bonding of the interface. In the case of $V_{O-4}$ vacancy formation, the Pt atom transfers 1.5 $e$ to six O atoms as well. However, the three O atoms (O-1, O-2, and O-3 labeled in Figure. 4) in the GB core region get more electrons than the others, indicating the tight bonding of Pt to O atoms in the GB region. As Pt has similar coordination environments with its surrounding O atoms when neutral $V_{O-2}$ or $V_{O-3}$ vacancies form, we present and discuss only the charge difference of Pt in the $V_{O-2}$-containing Pt@GB. In Fig. 7c, besides the charge accumulated on the four Pt-O bonds, we also find some electrons located on the Pt atom. This is consistent with the results of Bader-charge analysis. The Pt atom transfers only 0.78 $e$ to the four nearby O atoms, which is only about half the transferred charged in the other systems discussed. Since the neutral vacancy forming at the $V_{O-2}$ site makes Pt lose two nearest neighbor O atoms, the Pt atom just transfers a smaller number of electrons, leaving Pt with a local charge. As a result, the formation of a neutral $V_{O-2}$ or $V_{O-3}$ vacancy largely inhibits Pt-O bonding, and is disadvantageous for the segregation of Pt. However, the $V_{O-2}^{\bullet\bullet}$ vacancy formed at the GB has six nearest neighbor O atoms, and the Pt atom transfers 1.5 $e$ to the six O atoms. In this case the charge density difference (Fig. 7d) looks similar to that for the Pt@GB (Fig. 7a) and the Pt@GB-$V_{O-4}$ (Fig. 7e).

**Figure 7. Charge density differences are shown in two side views for several configurations at the GB. Part (a) is for the Pt@GB, while parts (b) to (e) are for**

the oxygen vacancy formed at the $V_{O-1}$, $V_{O-2}$ (neutral and 2+ charged), and $V_{O-4}$ sites, respectively. The charge accumulation and depletion are indicated by the yellow and blue regions, respectively. The isosurface value is 0.005 e/a.u.$^3$.

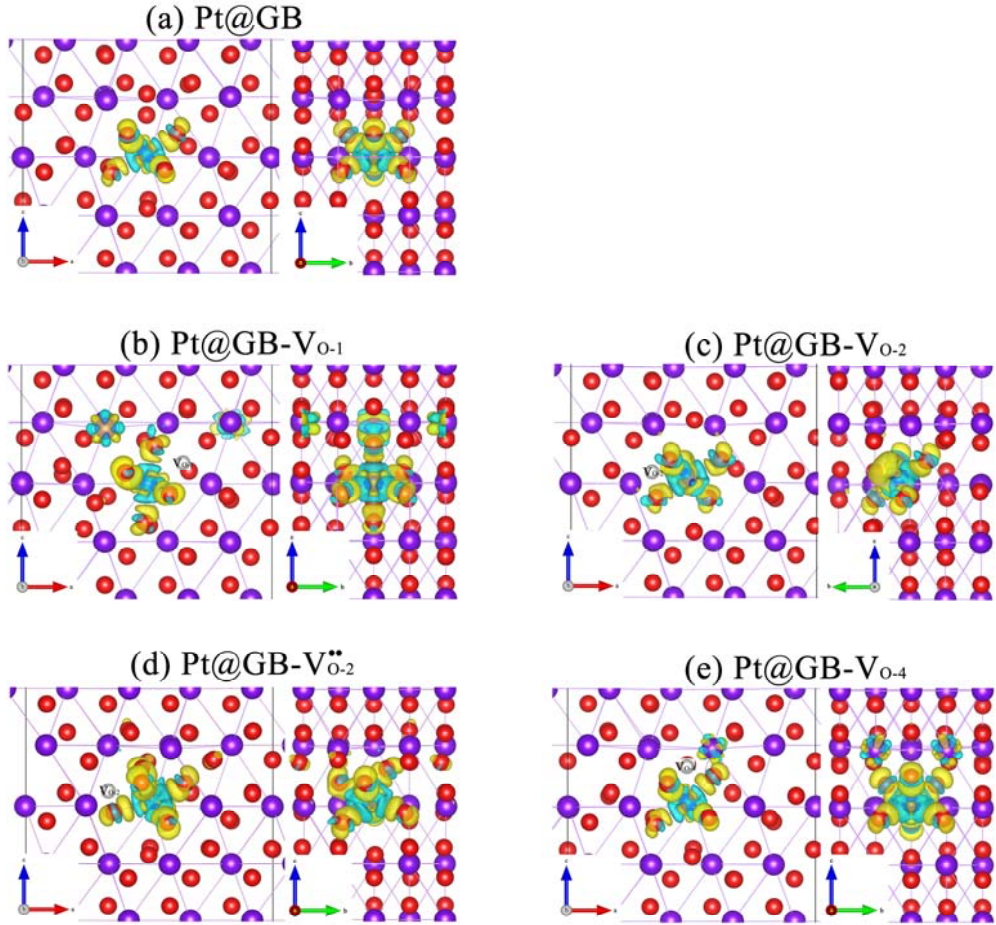

The magnetic properties induced by oxygen vacancies in the $CeO_2$ ∑3 (111) GB with and without a Pt atom are summarized in TABLE I. When an oxygen vacancy is present in the pristine GB, the total magnetic moment (TMM) is changed from zero (non-magnetic, NM) to 2.0 $\mu_B$ (ferromagnetic, FM), accompanied by a charge state transformation from 2+ to neutral. The spin density distribution shown in Fig. 8 shows that the magnetization is particularly localized at the Ce atoms which are the second nearest neighbors of the oxygen vacancy. When the oxygen vacancy is trapped at the GB-$V_{O-1}$ site next to the Pt atom, the FM configuration is favored, and the TMM is increased from 1 to 2 $\mu_B$. The spin density distribution is shared by two Ce atoms that are at the same distances from $V_{O-1}$. In the case of $V_{O-2}$ at the Pt@GB, no magnetic moment is observed, regardless of the charge state of the vacancy, and the system remains nonmagnetic as is the pristine $CeO_2$ GB. A nonmagnetic result is also obtained for the neutral state of $V_{O-3}$ which is less stable than the other three vacancies. When a vacancy forms at the $V_{O-4}$ site, the FM phase is dominant over almost the whole range of Fermi levels. The spin density distributes over the same Ce atoms as for the pristine Pt@GB.

**TABLE I.** The total magnetic moment (TMM) (in $\mu_B$), magnetic phase (MP), and spin-density distribution (The isosurface value is 0.005 e/a.u.$^3$) for the pristine $CeO_2$ $\sum 3$ (111) GB and for the Pt@GB with a vacancy.

| | Pristine GB | Pt@GB | | | |
|---|---|---|---|---|---|
| vacancy | $V_O^{\bullet\bullet}$ / $V_O^{\times}$ | $V_{O-1}^{\bullet\bullet}$ / $V_{O-1}^{\times}$ | $V_{O-2}^{\bullet\bullet}$ / $V_{O-2}^{\times}$ | $V_{O-3}^{\bullet}$ / $V_{O-3}^{\times}$ | $V_{O-4}^{\bullet\bullet}$ / $V_{O-4}^{\bullet}$ / $V_{O-4}^{\times}$ |
| TMM | 0.0/2.0 | 1.0/2.0 | 0.0/0.0 | 1.0/0.0 | 0.0/1.24/2.0 |
| MP | NM/FM | FM/FM | NM/NM | FM/NM | NM/FM |
| Spin-density | 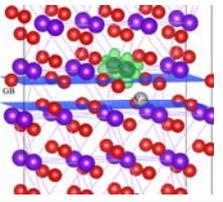 | 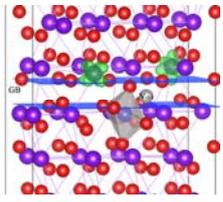 | - | unstable | 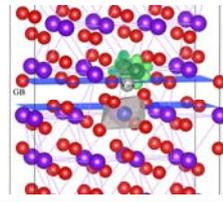 |

To provide insight into the strong interaction between a Pt atom and an oxygen vacancy, we used DOS analysis. Figure 8 shows the DOS for the vacancy formed at the most favorable $V_{O-1}$ site in the Pt@GB. The total DOS (TDOS) of the $V_{O-1}$-containing GB with and without a Pt atom are shown in panel (a). The two TDOSs look very similar, except that a bonding and an antibonding state appear separately in the Pt@GB, at the low and high energy levels, respectively. In the pristine GB-$V_{O-1}$, we can see a sharp peak in the spin-up state close to the Fermi level that may be considered as the defect state for $V_{O-1}$. To illustrate further details, the partial charge density for this defect state is shown in Fig. 8c as well. The additional charge for the neutral oxygen vacancy is equally distributed over the two next nearest

Ce atoms which are located on the other side of the GB. They each get one electron from the oxygen vacancy. In the Pt@GB-V$_{O-1}$ system, the defect state shifts downward away from the Fermi level, and is split into two sharp peaks at the spin-up state, corresponding to the partial charge density drawn in Fig. 9d. In this case, two Ce atoms obtain unequal numbers of electrons from the additional charge of the V$_O^\times$. The Ce atom farther from Pt gets more electrons than does the closer one, and thus the energy level (-0.49 eV) of the former is slightly lower than that of the latter (-0.42 eV). Obviously, the interaction between Pt and the oxygen vacancy lowers the level associated with the defect. In addition, the PDOS of the Pt atom in the vacancy-containing GB is also different from that in the GB without a vacancy. Comparing with the PDOS of Pt in the vacancy-free GB shown in Fig. 2b, there is an additional bonding state at the lowest energy level of -5.25 eV as shown in Fig. 9b, which is induced by strong hybridization between the Pt $5d_{3z^2}$ and the $2p$ states of the two O atoms above and below it. The corresponding partial charge visually exhibits the strong interaction of the Pt-O bond in the $z$ direction. Moreover, a sharp peak at the spin-up state is found at the -0.42 eV energy level in Fig. 9b, which is due to weak antibonding between the Pt $5d_{yz}$ and O $p$ states. Interestingly, one of the defect states, i.e., the Ce atom with fewer electrons due to its proximity to the Pt atom, occupies the identical energy level. Thus, the degeneration and reduction of the defect level is due to the fact that the segregated Pt atom shares only partial electrons with one Ce atom. Therefore, the formation of the vacancy changes the bonding direction of the Pt atom, and enhances the stabilization of Pt.

**Figure 8. (a) shows the total DOS for the vacancy-containing GB with (red line) and without (blue line) a segregated Pt atom. The PDOSs of the Pt atom and the O atoms above and below the Pt for the GB-V$_{O-1}$ system are presented in (b). The partial charge densities for four specific energy levels are shown in (c) to (f). Part (c) is for a defect state close to the Fermi level in the pristine GB. The partial charge corresponding to the Pt $5d_z^2$ and O $2p$ binding state at -5.25 eV is shown**

in panel (d). Parts (e) and (f) give the partial charges of the defect states in the Pt@GB-$V_{O-1}$ for the energy levels at -0.42 and 0.49 eV, respectively. The isosurface value is 0.003 e/ a.u.$^3$. The region between the two blue planes indicates the GB. The white ball represents the oxygen vacancy at the $V_{O-1}$ site.

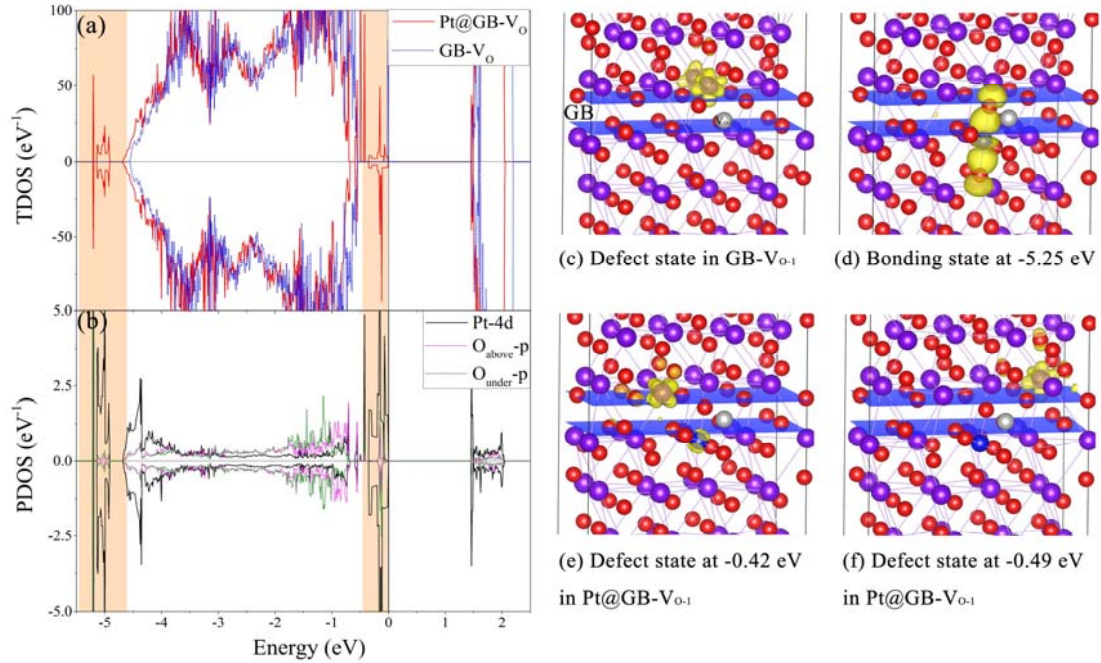

## SUMMARY

To summarize, we have carried out spin-polarized first-principles calculations to investigate the segregation tendency of Pt and the formation of oxygen vacancies at CeO$_2$ ∑3 (111) GBs.  Firstly, Pt has a very strong tendency to segregate to the CeO$_2$ ∑3 (111) GB because there are six nearest neighbor O atoms surrounding Pt, making Pt more stable at the GB than in the bulk. We have also calculated the solution energies of Pt in the bulk and on three kinds of CeO$_2$ surfaces. By comparing the structural stability of Pt in different configurations, it has been found that GBs are more favorable for Pt segregation than are the bulk or (111) and (110) surfaces, but are less favorable than the (100) surface. Lattice distortion plays an important role in the strong tendency of Pt to segregate. When Pt segregates to the GB, oxygen vacancies form spontaneously at the GB region, independent of environment conditions. In this process, although Pt loses one of its nearest O atoms, it finds a new neighbor on the other side of the GB, maintaining six nearest O atoms surrounding it.

However, without a segregated Pt atom, oxygen vacancies can spontaneously form only under O-poor conditions. As a result, Pt at the GB promotes the formation of oxygen vacancies, and their strong interaction enhances the binding of the interface. We propose that extra GBs fabricated close to the surfaces of nanocrystalline ceria can induce the segregation of Pt out of the bulk or other types of surfaces, resulting in the suppression of the accumulation and aggregation of Pt on the surface in the presence of redox reactions, especially under the O-poor conditions. In future research, we will further investigate the segregation of precious metals at perovskite GBs, to explore the remarkable potential of GBs in the self-regeneration of three-way catalysts.

**ACKNOWLEDGMENTS**

We would like to thank Dr. N. E. Davison for his kind help with the language. This paper is based on work supported by the National Natural Science Foundation of China (nos. 11347188 and 11404089), and the Natural Science Foundation of Hebei Province (no. A2015205142). The calculations were performed on the Quantum Materials Simulator of Hebtu.